\documentclass[aps,preprint]{revtex4}
\usepackage{hyperref}
\usepackage{amsfonts}
\usepackage{amsmath}

\usepackage{amssymb}
\usepackage{graphicx}%
\usepackage{bigints}
\usepackage{graphicx}
\usepackage{subfigure}
\usepackage{epsf}
\usepackage{bm}
\usepackage{amsmath}
\usepackage{amsfonts}
\usepackage{amssymb}
\usepackage{graphicx}
\usepackage{tabularx}
\usepackage{color}%
\providecommand{\U}[1]{\protect\rule{.1in}{.1in}}

\newcommand{\be}{\begin{equation}}
\newcommand{\ee}{\end{equation}}

\newcommand{\mincir}{\raise
-3.truept\hbox{\rlap{\hbox{$\sim$}}\raise4.truept\hbox{$<$}\ }}
\newcommand{\magcir}{\raise
-3.truept\hbox{\rlap{\hbox{$\sim$}}\raise4.truept\hbox{$>$}\ }}

\providecommand{\U}[1]{\protect\rule{.1in}{.1in}}

\usepackage{tikz,xcolor,hyperref}

\begin{document}

\title{Three dimensional charged black holes in Gauss-Bonnet gravity
}
\author{Kimet Jusufi$^{1}$}
\email{kimet.jusufi@unite.edu.mk}
\author{Mubasher Jamil$^{2}$}
\email{mjamil@sns.nust.edu.pk (Corresponding author)}
\author{Ahmad Sheykhi$^{3}$}
\email{asheykhi@shirazu.ac.ir}
\affiliation{$^{1}$Physics Department, State University of Tetovo, Ilinden Street nn, 1200, Tetovo, North Macedonia}
\affiliation{$^{2}$School of Natural Sciences, National University of Sciences and Technology, Islamabad, 44000 Pakistan,}
\affiliation{$^{3}$Physics
Department and Biruni Observatory, College of Sciences,\\
Shiraz University, Shiraz 71454, Iran}

\begin{abstract}
By using the zero-point length effect, we construct a new class of  charged black hole solutions in the framework of three
dimensional Gauss-Bonnet (GB) gravity with Maxwell electrodynamics. The gravitational and
electromagnetic potentials are
finite and regular everywhere, however, the computation of  spacetime scalar curvature invariants suggest the presence of a singularity at the origin. We also explore thermodynamics of
the obtained solutions and reveal that the entropy of the black
hole decreases due to the stringy effects. The
thermodynamic and conserved quantities are computed and also the validity of
the first law of thermodynamics on the black hole horizon is verified.
Finally, the spinning black hole solution is also reported.
\end{abstract}
\maketitle

\section{Introduction}
Black holes which were predicted a century ago, are among the most fascinating scientific discoveries
whose existence has now been confirmed through detection of
the gravitational waves as well as direct image of their shadow, in recent years \cite{ligo,sgr,m87}.
Black holes are entirely geometrical objects but possess other enriched physical structures which connect
different branches of physics including quantum gravity,
thermodynamics, superconducting phase transition,
paramagnetism-ferromagnetism phase transition, superfluids,
condensed matter physics, spectroscopy, information theory,
holographic hypothesis, etc. A massive black hole is usually
formed by a relentless gravitational collapse when a supermassive star collapses at the final stage of its life, for review see \cite{joshi}, however the outcome of collapse could be a naked singularity as well.
According to the singularity theorems of Penrose and Hawking \cite{pen}, all the mass constituting the 
star eventually lumps to a singular point and further during the collapse
a geometrical surface called horizon is formed which hides the
singularity from viewing for all observers away from the black hole. It is a general belief
that the singularities are nonphysical objects which are created
by classical theories of gravity (ignoring quantum mechanical effects which would otherwise dominate over gravity near the Planck length scale) and they do not exist in nature. In literature, attempts have been made to study a black hole and a naked singularity alongside and differentiate them using different astronomical observations \cite{jamil}.
Regular black holes are those that avoid singularity at their
center \cite{Bar1,Mars,Bor}. In particular the Bardeen solution \cite{Ayon-Beato:2000mjt} was letter found to be an exact solution by incorporating a magnetic charge using non-linear electrodynamics. However, some of the proposed black holes are not the solutions of Einstein field equations,
since the physical sources producing them are unknown. The best
candidate today to produce singularity-free solutions, even at the
classical level, due to its intrinsic non-locality, is string
theory \cite{Ts}. Regular black holes have been also studied in the context
of supergravity \cite{Cv}, conformal massive gravity \cite{conformal}, renormalization group approach \cite{group} as well as conformal field theory
\cite{Hor}. A general procedure for constructing exact black hole
solutions with electric or magnetic charges in Einstein gravity is
presented in \cite{Zh}. It was shown that in order to reproduce a
regular black hole, one needs to take into account the nonlinear
electrodynamics as the gauge field \cite{Zh}. Other investigations
on the regular black hole solutions are carried out in
\cite{Ayon1,Ayon2,Bog,Hay,Bam,Gh,Tos}.

The discovery of three dimensional solutions of general relativity
in anti-de Sitter (AdS) spacetime known as 
Banados-Teitelboim-Zanelli (BTZ) black hole  \cite{BTZ1} and Martinez-Teitelboim-Zanelli (MTZ) \cite{mtz}, has been among
the notable achievements in black hole physics. Later, these three dimensional black holes were generalized via Noether symmetry approach in $f(R)$ gravity \cite{darabi}. In fact, the
$(2+1)$-dimensional solution of Einstein gravity provides a toy
model to investigate and find some conceptual issues such as black
hole thermodynamics, quantum gravity, string and gauge/gravity
duality, holographic superconductors in the context of the
$AdS_{3}/CFT_{2}$ \cite{Car1,setare,Ash,Sar,Wit1,Car2}. It has been shown
that the quasinormal modes in this spacetime coincide with the
poles of the correlation function in the dual CFT. This gives
quantitative evidence for $AdS_{3}/CFT_{2}$ \cite{Bir}.
Furthermore, BTZ black holes play a crucial role for improving our
perception of gravitational interaction in low dimensional
spacetimes \cite{Wit2}. In particular, it might shed some light on
the quantum gravity in three dimensions. It was shown that the
surface $r=0$ is not a curvature singularity but, rather, a
singularity in the causal structure which is everywhere constant
and continuing beyond it would produce closed timelike curves
\cite{BTZ2}. The extension to include an electric charge in
addition to the mass and angular momentum have been performed
\cite{BTZ3,Fern}. The $(2+1)$-dimensional black holes also provide
a powerful background to explore one-dimensional holographic
superconductors \cite{Ren, Liu, Kord,mahya,Bina}. In three
dimensional $f(R)$ gravity with a self-interacting scalar field
non-minimally coupled to gravity, the hairy black hole solutions
have been already derived and shown to be thermodynamically stable
\cite{wang}. Black holes containing magnetic charge and
non-linearity parameter in Born-Infeld gravity have been
constructed in  \cite{ali}. As a gravitational analogue, three
dimensional wormholes have also been derived and investigated
thermodynamically \cite{Jamil1}. In lower spatial dimensions, the
spacetime topologies are simpler which facilitate to investigate
the classical and quantum properties of black holes and
wormholes in details \cite{angel}. Additional studies on
$(2+1)$-dimensional black hole solutions of gravitational field equations
have been extensively carried out in the literature (see e.g.
\cite{rin1,rin2,Clem,Car,Noj3D,Emp,Cad,Par,Hendi,She,Xu,Grum2,
Shey3D}).

In the present work we are going to investigate regular black hole
solutions in $(2+1)$-dimensional GB gravity by using zero-point
length effect. The study on the lower dimensional black hole
solutions of GB gravity have received a lot of attentions in
recent years \cite{HKM1,HKM2,Mel}. These solutions indeed
generalize the BTZ black hole from Einstein gravity to GB gravity, and restore the three dimensional BTZ
solutions of Einstein gravity when the GB coupling goes to zero.

This article is structured as follows. In the section II, we
construct three dimensional black hole solutions of GB gravity with finite
electrodynamics. In section III, we explore thermodynamics of three dimensional
 black holes in GB gravity. In section IV, we express the Einstein field equation as a thermodynamical relation.  The last section is devoted to
conclusion and discussion. All Greek indices vary from $0,1,2$. Units are chosen such that $c=1=G$.
\section{3D black holes in GB gravity with Maxwell electrodynamics}
Using the regularization factor $r \to \sqrt{r^2+l_0^2}$, where
$l_0$ can play the role of zero point length
\cite{Jusufi:2022ukt}, for the gravitational potential by means of
regularization factor in $(2+1)$ dimensions it was shown that
\cite{Jusufi:2022nru}
\begin{equation}
    \Phi(r)=k M \ln\left( \sqrt{r^2+l_0^2}\right),\label{potential}
\end{equation}
where $\Phi(r)$ is the gravitational potential per unit mass and $k$ is a constant of proportionality. By drawing upon concepts from string T-duality, it has been demonstrated that the zero-point length, denoted as $l_0$, assumes the role akin to that of the Planck length (see \cite{l1,l2,l3,l4}). From the form of the last equation we observe that one has units of length inside the logarithm, however, one can add a normalization constant $r_0$ inside logarithmic terms, hence in general we have $\ln\left( \sqrt{r^2+l_0^2}/r_0\right) $. However, this constant will be omitted in all logarithmic terms in our paper. Using the gravitational potential mentioned above, we can solve the
Poisson's equation in polar coordinates  
\begin{equation}
   \nabla^2 \Phi(r)=2 \pi \rho,
\end{equation}
which easily yields the energy density as follows \cite{Jusufi:2022nru}
\begin{equation}
\rho(r)=\frac{k\,M\,l_0^2}{\pi (r^2+l_0^2)^2}.
\end{equation}
We can already see that when $l_0\to 0$, $\rho(r)$ reduces to zero, giving a
point mass distribution. In contrast, here due to the zero point
length, we get a smeared matter distribution due to quantum
gravity effects described by the quantum modified energy momentum
tensor ${(T^\mu}_\nu)^{str} = \it{diag}
\left(-\rho,p_r,p_{\varphi}\right)$. It is interesting to note that this energy density can be reduced to the energy density given in ref.\cite{Estrada:2020tbz} that also leads
to a regular solution in Einstein gravity. For the electromagnetic
potential in $(2+1)$-dimensions we have \cite{Jusufi:2022nru}
\begin{equation}\label{Amu}
A_\mu=(-Q \ln(\sqrt{r^2+l_0^2}),0,0),
\end{equation}
where $A_{\mu}$ is the three-potential which is well-defined even when $r \to 0$, for these reasons, we can refer to such a theory as a finite electrodynamics which is basically the Maxwell electrodynamics with a regular potential in the limit $r \to 0$.
By comparing (\ref{potential}) and $A_t$ in (\ref{Amu}), we notice that $Q=-kM$.  This has to do with the fact that both the gravitational force and electrostatic force for example in 4D have a similar law. In 3D, it is no surprise to see that potentials are similar up to a sign difference of course due to the nature of charge.
A regularized GB theory has been shown to
exist and it belongs to what is known as the scalar-tensor
formulation described by the action
\cite{Hennigar:2020lsl,Fernandes1,Fernandes2}:
\begin{equation}
S= \int_{\mathcal{M}} d^{D} x \sqrt{-g}\Big{\{}R-2\Lambda+\alpha \big[4 G^{\mu \nu} \nabla_{\mu} \phi \nabla_{\nu} \phi+\phi \mathcal{G}+4 \square \phi(\nabla \phi)^{2}+2(\nabla \phi)^{4}\big]
\Big{\}}+S_{EM}+S_{M},  \label{action}
\end{equation}
where $\phi(x^\mu)$ is a real scalar field, $S_{EM}$ is the action for the electromagnetic field, $S_{M}$ is the matter action which arises due to the quantum gravity effects encoded in $l_0$, finally $\mathcal{G}=R_{\mu
\nu\lambda\rho}R^{\mu \nu\lambda\rho}-4R_{\mu \nu}R^{\mu \nu}+R^2$
is the GB term. In fact,
the action in the last equation is shown to be a special case of a
more general well-defined action given in \cite{Lu:2020iav}
\begin{eqnarray}\notag
S&=&\int d^D x\sqrt{-g}\Big{\{}R
-2\Lambda+\alpha\Big(\phi\,\mathcal{G}+4G^{\mu\nu}\partial_\mu\phi\partial_\nu\phi
-2\eta Re^{-2\phi}-4(\partial\phi)^2\Box\phi+2((\partial\phi)^2)^2\\
&-&12\eta(\partial\phi)^2e^{-2\phi}-6\eta^2e^{-4\phi}\Big)\Big{\}}\,.
\end{eqnarray}
Here $G^{\mu \nu}$ denotes the Einstein tensor. In this approach, one can use the Kaluza-Klein-like procedure to
generate the limit of GB gravity by means of compactification.
This action depends on a parameter $\eta$ which describes the
curvature of the (maximally symmetric) internal space
\cite{Lu:2020iav}
\begin{equation}
R_{\mu \nu\lambda\rho}= \eta(g_{\mu \lambda}g_{\nu
\rho}-g_{\mu\rho}g_{\nu\lambda})\,.
\end{equation}
The equivalence is achieved for vanishing $\eta$. As was argued
in \cite{HKM1,HKM2}, there is no logical obstruction to consider a
lower-dimensional GB gravity by setting $D = 3$ in the above
action to obtain a $D = 3$ version of GB gravity. Note that $\mathcal{G}$ vanishes in $D = 3$  by construction or the choice of the metric. Also note that we are working with a regularized theory or a scalar-tensor theory of gravity. Vanishing $\mathcal{G}$ in the action (6) simplifies the computations but we still have an effect of $\alpha$. In what follows
we shall consider this theory where $\mathcal{G}=0$ for $D
= 3$. Following \cite{HKM1,HKM2}, we can consider a static
solution with
\begin{equation}
    ds^2=-f_{GB}(r)dt^2+\frac{dr^2}{f_{GB}(r) h(r)}+r^2 d\varphi^2.
\end{equation}
For the matter terms we have the $S_{M}$ which encodes the effect of string corrections (terms containing zero-point length) given by
\begin{eqnarray}
   S_M=\int d^3x \sqrt{-g}\,\,\mathcal{L}_M.
\end{eqnarray}
The simplest choice is to take $\mathcal{L}_M \sim \rho$.
 We do not have yet the full Lagrangian regarding the quantum gravity effects. In this direction it is also possible that the Einstein field equations might be modified. However, having the energy density we can obtain the effective solution (along with effective Einstein field equation) and one way of doing it is to identify $\mathcal {L}_M$ with the scalar $\rho$.

If we solve (2) in the background metric (8) we get
\begin{eqnarray}
\mathcal{L}_M=\frac{k M l_0^2 h(r)}{\pi (r^2+l_0^2)^2}+\frac{k h'(r) M r}{4 \pi (r^2+l_0^2)},
\end{eqnarray}
which $\mathcal{L}_M$ encodes the smeared matter distribution due
to stringy effects. The potential (\ref{potential}) corresponds to
the virtual-particle exchange obtained in flat spacetime and, in general, that means we can set $f(r)\to 1$ in metric (8). Although this is a good approximation, and can be justified from the equivalence principle $f(r)\to 1$, in our case, however, we have a nontrivial metric due to the presence of $h(r)$. With this information in hand, we solve $\nabla^2\Phi=2 \pi \rho$ and as we see from the last equation, it contains (3) in the special case by setting $h(r)=1$ and $h'(r)=0$. The reason why we need this relations will be clear later on since we need the equations of motions where we have to take a variation w.r.t $h$.. The
finite electrodynamics is described by the action
\begin{eqnarray}
   S_{EM}=- \int d^3x \sqrt{-g}\mathcal{L}_{EM},
\end{eqnarray}
and one can evaluate the Faraday tensor $F_{\mu \nu}$ using Eq. (4) and then get the Lagrangian for the electromagnetic field given by \cite{Jusufi:2022nru}
\begin{equation}
\mathcal{L}_{EM}=\sigma F^{\mu \nu}F_{\mu \nu}= \sigma\,\frac{2\,Q^2\, r^2\,h(r)}{(r^2+l_0^2)^2},
\end{equation}
where $\sigma$ is a constant of proportionality and to be constrained via astronomical observations.
Having in mind that the GB term vanishes in three dimensions, i.e.
$\mathcal{G}=0$, and working with a vanishing $\eta$, for the
total effective Lagrangian density we have
\begin{eqnarray}
\mathcal{L}_{tot}=R-2\Lambda+\alpha \big(4 G^{\mu \nu} \nabla_{\mu} \phi \nabla_{\nu} \phi
+ 4 \square \phi(\nabla \phi)^{2}+2(\nabla \phi)^{4}\big)+\mathcal{L}_{EM}+\mathcal{L}_{M}.
\end{eqnarray}
Following \cite{HKM1,HKM2}, one has to use the Euler-Lagrange
equations and can obtain
three equations of motion for the variables $f_{GB}, h, \phi$.
It is worth noting
that the equation of motion for the above Lagrangian keep the
fundamental properties of GR and its extensions. As a result the
equation of motion of second order respect to the metric and
energy momentum tensor are conserved.
In particular, focusing on the BTZ
solutions that satisfy the condition $h = 1$, and fixing
$\sigma=1/2$, the equation coming from the variation w.r.t
$f_{GB}$ reads
\begin{eqnarray}
\alpha \left(\phi'^2+\phi''\right)\left(r\phi'-1\right)=0,
\end{eqnarray}
which has the solution $\phi(r)=\ln(r/l)$, with $l$ being a constant of integration. On the other hand, the equation coming from variation  w.r.t $h$ yields
\begin{eqnarray}
\frac{f_{GB}'(r)}{2r}+\frac{\alpha f_{GB}(r) f_{GB}'(r)}{r^3}-\frac{\alpha f_{GB}(r)^2}{r^4}+\frac{M l_0^2 }{(r^2+l_0^2)^2}
+\frac{ Q^2r^2}{(r^2+l_0^2)^2}-\frac{1}{l^2}=0.
\end{eqnarray}
Solving the above equation and identifying the constant of integration with $M$, we find the following metric coefficient

\begin{equation}\label{fgb}
    f_{GB}(r)=-\frac{r^2}{2\alpha}\left(1 \pm \sqrt{1+\frac{4\alpha}{r^2}f_E(r)}\right),
\end{equation}
where
\begin{equation}
    f_E(r)\equiv\frac{r^2}{l^2}-\frac{M r^2}{r^2+l_0^2}-\frac{ Q^2l_0^2}{r^2+l_0^2}- Q^2\ln(r^2+l_0^2).
\end{equation}
Notice an interesting property of the above solution: if we use the transformation $r \to -r$, that does not generate a solution with negative mass but a replica of the same spacetime geometry. This fact can be seen also in the expression for the energy density in Eq. (3). In Fig. 1 (left hand side), we have shown a plot of the function $f_{GB}(r)$ from where one can see the presence of the event horizon for this particular black hole and the plot for $\rho(r)$ (right hand side). Another important aspect of our black hole solution is that it diverges at asymptote infinity. Namely, this can be seen from the presence of the first term $r^2/l^2$, but also the presence of last term $ Q^2\ln(r^2+l_0^2)$ in Eq. (17). Eliminating these divergences (logarithmic and $r^2/l^2$ divergences), is beyond the scope of the present paper, however, one possibility is by using counter-terms (see in particular \cite{Aros:1999kt}). But again, this requires further investigation of our solution. Usually in Gauss-Bonnet gravity we get two solutions namely (\ref{fgb}) which differ from the sign. The $+$ branch is not a physical solution in general, for example we can have a $+$ sign before the mass term  which signals a diverging mass for large radii. Moreover the behavior of the core can be obtained if we analyze $r<l_0$, it turns out that core can behave as a de-Sitter metric. This fact was explained in \cite{Jusufi:2022nru}.

\begin{figure*}[h!]
 \includegraphics[width=7.5cm,height=7.5cm]{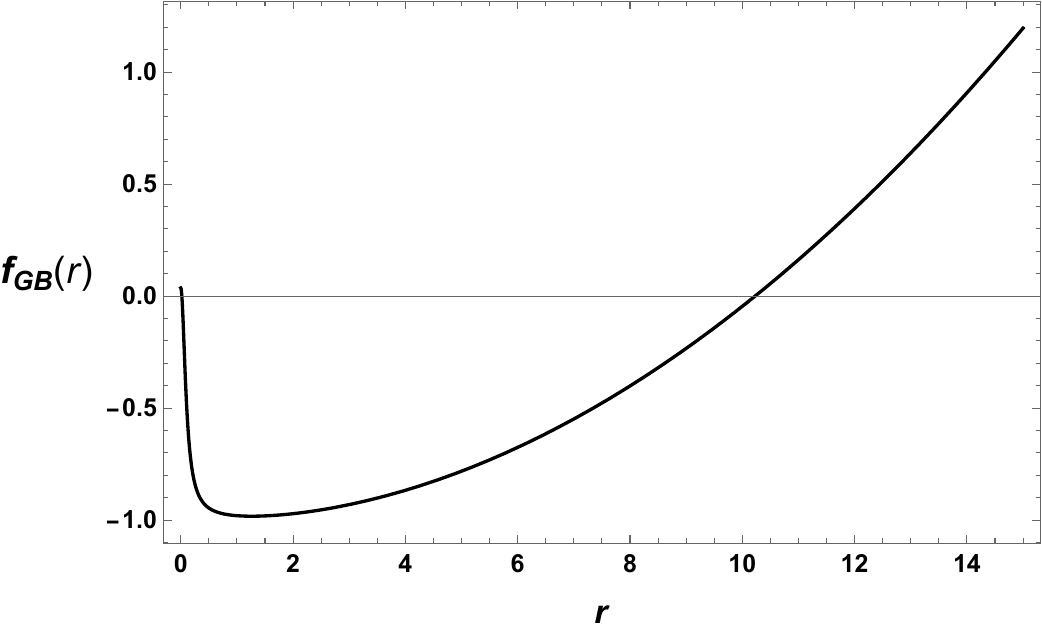}
\includegraphics[width=7.5cm,height=7.5cm]{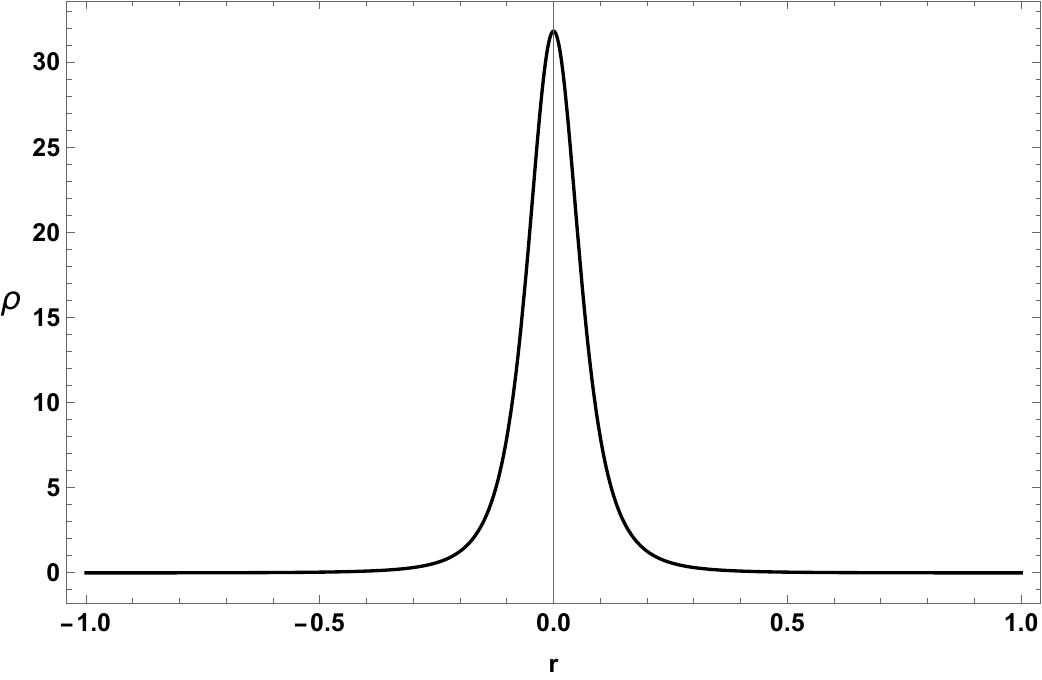}
  \caption{Left panel: Plot of $f_{GB}(r)$ as a function of $r$. We have set $M=1$, $Q=0.1$, $l=10$, $l_0=0.1$ and $\alpha=0.001$. This particular case shows the presence of the horizon, hence the solution describes a black holes. Right panel: Plot of the energy density $\rho$ (given by Eq. (3)) as a function of $r$. We have set $M=1$ and $l_0=0.1$.}
\end{figure*}

The above result is true provided the following conditions hold:
(i) Consider the scalar-tensor limit of GB gravity like
\cite{HKM1,HKM2} coupled to additional matter, (ii) The matter
considered has the property that $T_t^t = T_r^r$. (iii)
Considering three-dimensional generalizations of the BTZ black
hole, then (16) is true. Basically, it just needs to be the case
that there is a single field equation for the gravitational
sector. This was noted in \cite{HKM1}, one expects the form of Eq.
(16) to remains valid for charged GB BTZ black holes even in any
theory of nonlinear electrodynamics characterized by the action.
The negative sign branch of the above admits a well-defined limit
as $\alpha \to 0$, yielding
\begin{equation}
    f_{GB}(r)=f_E-\frac{\alpha}{r^2}f_E^2+\dots
\end{equation}
in agreement with \cite{Jusufi:2022nru} in leading order terms. We
can see that in the limit of $l_0\to 0$, the charged black hole
solution in 3D GB gravity, reported in Ref. \cite{HKM1} is
obtained
\begin{equation}
    f_E(r)=\frac{r^2}{l^2}-M-2Q^2\ln(r).
\end{equation}
At the horizon we need $f_{GB}(r_+)=0$. This means $1-\sqrt{1+4 \alpha f_E(r_+)/r_+^2}=0$, yielding $f_E(r_+)=0$ but so does $f_{GB}(r_+)=0$.
 It is interesting to note here that the solution presented in \cite{Jusufi:2022nru} is free from singularities, the solution (16) in the present paper is not, namely one can show that the curvature scalar invariants such as the Ricci scalar and Kretschmann scalar diverge in the limit $r \to 0$, namely $R=R_{\mu \nu} g^{\mu \nu}$ and $K=R_{\mu \nu \sigma \rho} R^{\mu \nu \sigma \rho}$, blows up to $-\infty$ and $+\infty$, respectively. We have shown the plot of Ricci scalar and Kretschmann scalar in Fig. 2. In addition, the curvature is present also for $\alpha<0$. This follows from 
 the fact that the Kretschmann is function of the derivative of \ref{fgb} with the  inner-square term $1+4 \alpha f_E(r)/r$ should have a zero (a curvature singularity), i.e the solution should have a zero also for $\alpha<0$.

\begin{figure*}[h!]
 \includegraphics[width=7cm,height=7cm]{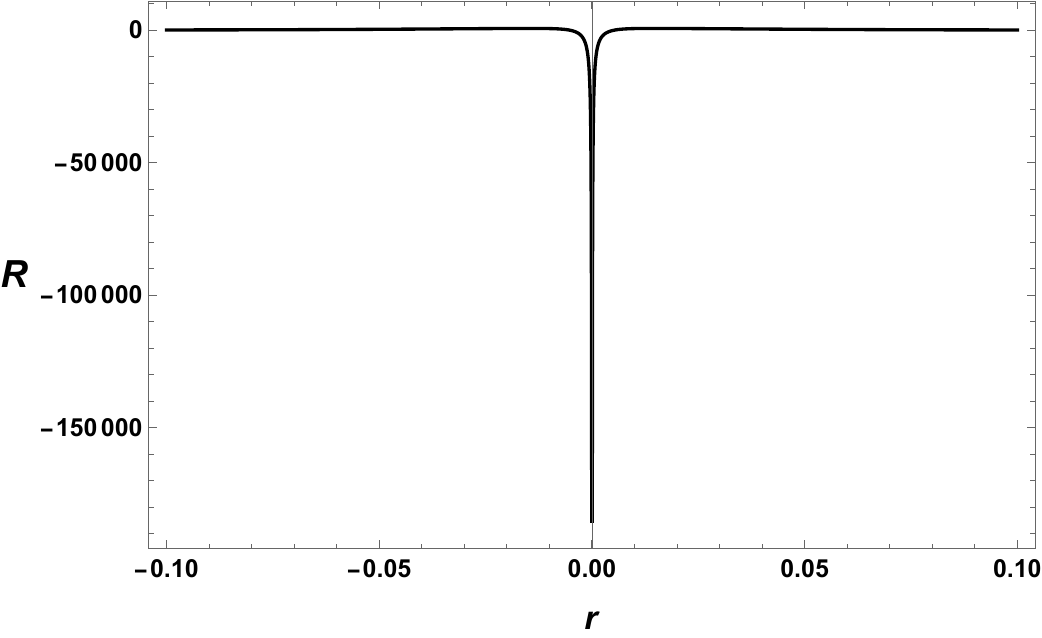}
  \includegraphics[width=7cm,height=7cm]{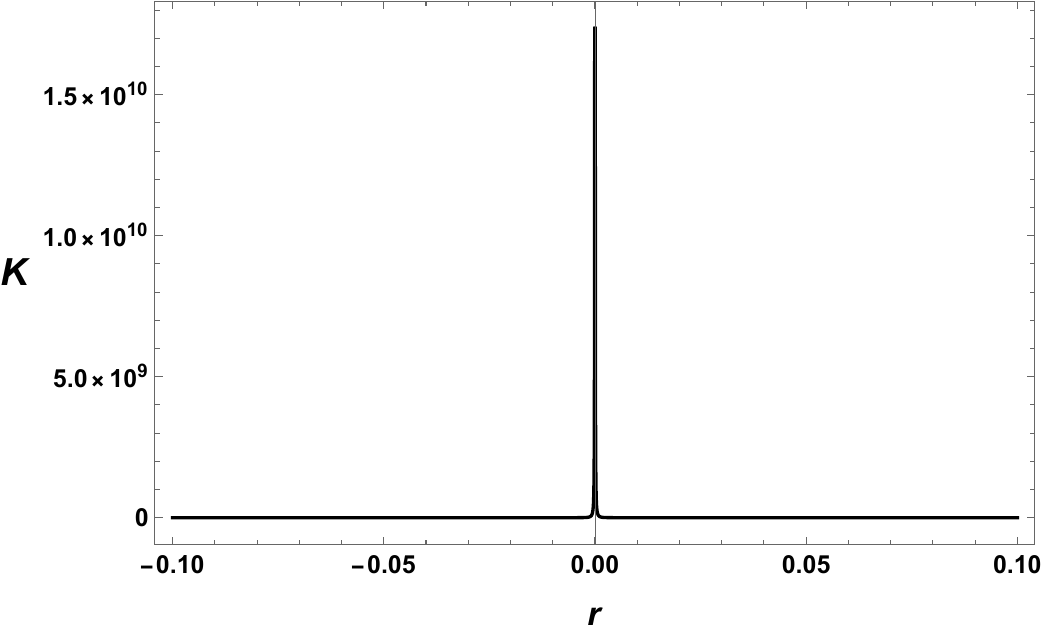}
  \caption{Plot of Ricci scalar and Kretschmann scalar. We have set $M=1$, $Q=0.1$, $l=10$, $l_0=0.1$ and $\alpha=0.001$. This particular case shows the presence singularities.}
\end{figure*}

\section{Thermodynamics of 3D GB black holes}
In this section we shall focus on the thermodynamics of the black
hole solution (16). One can easily see that 3D charged GB black
holes share the same horizon radius $r_+$ and the same
temperature as in 3D general relativity, hence we can write
\begin{eqnarray}
T=\frac{f'(r)_{GB}}{4 \pi}|_{r_+}=\frac{f'(r)_{E}}{4 \pi}|_{r_+}
\end{eqnarray}
This means that their BH thermodynamics will be identical.  Following
\cite{HKM1,HKM2}, first we can define
\begin{eqnarray}
\mathcal{M}=\frac{M}{8},
\end{eqnarray}
where the mass parameter is obtained from $f_E(r_+)=0$,
\begin{eqnarray}
M=\frac{r_+^2}{l^2}+\frac{l_0^2}{l^2}-\frac{Q^2 l_0^2}{r_+^2}-\frac{Q^2 (r_+^2+l_0^2)\ln(r_+^2+l_0^2)}{r_+^2}.
\end{eqnarray}
Using the mass, it follows for the black hole temperature that
\begin{equation}
T=\frac{r_+}{2 \pi l^2}+\frac{ l_0^2 Q^2 \ln(r_+^2+l_0^2)}{ 2 \pi r_+ (r_+^2+l_0^2) }-\frac{ Q^2 (r_+^2-l_0^2)}{2 \pi r_+(r_+^2+l_0^2)}-\frac{ l_0^2 r_+}{2 \pi (r_+^2+l_0^2) l^2}.
\end{equation}
From Eq. (22) we also see that
\begin{equation}
    d\mathcal{M}=\bigg(\frac{\partial \mathcal{M}}{\partial r_{+} }\bigg)_{Q,l_0}dr_{+}+\bigg(\frac{\partial \mathcal{M}}{\partial Q }\bigg)_{{r_+},l_0}dQ+\bigg(\frac{\partial \mathcal{M}}{\partial l_0 }\bigg)_{r_{+},Q}dl_0,
    \end{equation}
where
    \begin{eqnarray}\notag
    &&\bigg(\frac{\partial \mathcal{M}}{\partial r_{+} }\bigg)_{Q,l_0}=\frac{ r_+}{4l^2}+\frac{ l_0^2 Q^2 \ln(r_+^2+l_0^2)}{4r_+^3}-\frac{ Q^2 (r_+^2-l_0^2)}{4r_+^3},\\\notag
    &&\bigg(\frac{\partial \mathcal{M}}{\partial Q }\bigg)_{r_{+},l_0}=-\frac{ (l_0^2+r_+^2) Q \ln(r_+^2+l_0^2)}{4\,r_+^2}=\phi_e ,\\\notag
    &&\bigg(\frac{\partial \mathcal{M}}{\partial l_0 }\bigg)_{r_{+},Q}=\frac{l_0}{4l^2}-\frac{ Q^2 l_0}{4r_+^2}-\frac{ l_0 Q^2 \ln(r_+^2+l_0^2)}{4r_+^2 }=\psi_{l_0},
    \end{eqnarray}
where $\phi_e$ and $\psi_{l_0}$ are the potentials conjugate to $Q$ and $l_0$, respectively. By using the fact that $d\mathcal{M}=T dS$ (keeping $Q$ and $l_0$ as constant), for the entropy we have
\begin{eqnarray}
S&=&  \int  \frac{1}{T} \bigg(\frac{\partial \mathcal{M}}{\partial r_{+} }\bigg)_{Q,l_0} dr_+.
\end{eqnarray}

Evaluating the last integral yields
\begin{eqnarray}\label{26}
S=\frac{\pi r_+}{2}\left(1-\frac{l_0^2}{r_+^2}\right),
\end{eqnarray}
where the constant of integration is set to zero.

From the last equation we see that the stringy effect suggests a
decrease in black hole entropy. 
In the limit $l_0\to 0$, we
get $\lim_{l_0 \to 0}S=\pi r_+/2$ \cite{HKM1,HKM2}. The presence of $l_0$ modifies the energy momentum tensor and modifies 
the first law of thermodynamics, and the
first law leads to entropy values that do not follow the standard area law. This was the case also for regular black hole studied in Ref. \cite{Ma:2014qma}. The surface
area of the black hole is
\begin{eqnarray}
A=\int_0^{2 \pi} r_+ d\varphi =2 \pi r_+,
\end{eqnarray}
hence we see that for the classical case with no stringy effect the Hawking-Bekenstein relation holds,
\begin{eqnarray}
S=\frac{A}{4}.
\end{eqnarray}
Let us note here that in the extended phase space, where the
cosmological constant is interpreted as a thermal pressure,  the
mass of the black hole is no longer regarded as internal energy,
rather it is identified with the chemical enthalpy \cite{mann}.
Adopting this picture, one can write $\mathcal{M}=E+PV$, with the
potential given by
\begin{eqnarray}
P&=& \frac{1}{8 \pi l^2},\,\,V=\pi r_+^2,
\end{eqnarray}
we can write the first law for the black hole,
\begin{eqnarray}
d\mathcal{M}=TdS+VdP+\phi_e dQ+\psi_{l_0} dl_0.
\end{eqnarray}
\section{Thermodynamic interpretation of the field equations}
Thermodynamics-Gravity conjecture has been well established
through numerus investigations \cite{Jac,Pad1,Pad2}. It has been
confirmed that one can always rewrite the field equations of
gravity in the form of the first law of thermodynamics. This
connection has been confirmed for Einstein, GB, and more general
Lovelock gravity \cite{Pad3,CaiKim,Cai2,SheT1}. For topological
gravity, it has been shown that the gravitational field equations
of $(n+1)$-dimensional topological black holes with constant
horizon curvature, in cubic and quartic quasi-topological gravity,
can be recast in the form of the first law of thermodynamics,
$dE=TdS-PdV$, at the black hole horizon \cite{Sheqt}. This
connection was also shown in the framework of brane cosmology,
where it was proved that the Friedmann equations describing the
evolution of the universe can be written in the form of the first
law of themodynamics on the apparent horizon and vice versa
\cite{Shey1,Shey2}.

Here we would like to examine the connection between the field
equations and first law of thermodynamics for regular GB black
holes in three dimensions. To this end, let us rewrite the
$(rr)$-component of the field equations in the presence of the
cosmological constant, at the horizon with the assumption that equation holds also for $r=r_+$, we then obtain as follows
\begin{equation}
\frac{f_{GB}'(r_+)}{2r_+}+\frac{\alpha f_{GB}(r_+)
f_{GB}'(r_+)}{r_+^3}-\frac{\alpha
f_{GB}(r_+)^2}{r_+^4}-\frac{1}{l^2}=8 \pi T^r_r|_{r_+}
\end{equation}
where $T^r_r=P_r$ is the radial pressure of matter at the horizon \cite{Pad3}, in our case, it is given by
\begin{equation}
   P_r|_{r_+}= - \frac{M l_0^2 }{8 \pi (r_+^2+l_0^2)^2}-\frac{ Q^2r_+^2}{8 \pi (r_+^2+l_0^2)^2}.
\end{equation}


Using the fact that at the horizon we need $f_{GB}(r_+)=0$, this means $1=\sqrt{1+4 \alpha f_E(r_+)/r_+^2}=0$, yielding $f_E(r_+)=0$ but so does $f_{GB}(r_+)=0$.
This implies that the second and third terms in the above equation
vanish. Moreover, at the horizon, as we saw
$f_{GB}(r_+)=f_{E}(r_+)$, this gives
\begin{equation}
\frac{f_{E}'(r_+)}{2r_+}-\frac{1}{l^2}=8 \pi P_r,
\end{equation}
which can also be rewritten as follows
\begin{equation}
\frac{f_{E}'(r_+)}{4 \pi}d\left(\frac{\pi
r_+}{2}\right)-\frac{d\left(r_+^2\right)}{8\,l^2}=P_r d(\pi
r_+^2).
\end{equation}
The second term on left side of the above equation is the change of mass, by using Eqs. (22-23), we can write
\begin{equation}
   d\left( \frac{r_+^2}{8l^2}\right)=d \mathcal{M}-d \left(\frac{l_0^2}{8 l^2}\right)+d\left(\frac{Q^2 (r_+^2+l_0^2)\ln(r_+^2+l_0^2)}{8r_+^2}\right)
\end{equation}
\begin{equation}
\frac{f_{E}'(r_+)}{4 \pi}dS+\phi_e dQ+\psi_{l_0}dl_0-d\mathcal{M}=P_r dV,
\end{equation}
where $\phi_e$ and $\psi_{l_0}$ are the potentials conjugate to $Q$ and $l_0$, respectively.
Finally, we can express the total mass of the system as follows
\begin{equation}
d\mathcal{M}=TdS+\phi_e dQ+\psi_{l_0}dl_0-P_rdV,
\end{equation}
where $dV$ is the change in horizon volume and the term $P_r dV$
corresponds to work done against the pressure. Note that this equation is obtained
from a combination of the equations of motion at horizon and the variation of parameters at horizon. Hence, the
field equations near horizon of the 3D GB black hole can be
expressed as a thermodynamic identity under the virtual
displacement of the horizon. The effect of zero-point length is encoded in the third term. On the other hand, there is no effect of the GB parameter $\alpha$, this is due to the fact that the solutions given by Eqs. (16-17) share the same horizon as a consequence the thermodynamical properties are the same. 
\section{Conclusion and discussion}
In a framework of 3D regularized GB theory of gravity which
belongs to a scalar-tensor formulation of gravity, we obtained an
exact black hole solutions using the zero-point length effect. The
black hole solution is described by the mass, electric charge and
the zero-point length $l_0$. The gravitational and electromagnetic
potentials are finite and
regular in the limit $r \to 0$, however, in general, the scalar curvature invaraints shows the presence of singularities at the origin. We investigated the thermodynamics
of the black hole solution, in particular we obtained the first
law for the black hole thermodynamics relation. Using
Thermodynamics-Gravity conjecture, we have shown that the field
equations of 3D GB black hole can be expressed as the first law of
thermodynamics on the horizon. Due to the stringy effects, we
found that the entropy of the black hole decreases. Finally, we
briefly investigated 3D rotating solution with zero-point length
effect in GB gravity.

  \section*{Appendix: Rotating black hole solutions}
 For the rotating black hole in three dimensions it was found \cite{Jusufi:2022nru}
 \begin{eqnarray} ds^2=
- N^2 dt^2 + r^2 \left(N^\varphi dt + d\varphi\right)^2 +
\frac{dr^2}{\hat{f}_E}, \end{eqnarray}

with
\begin{eqnarray}
N^2=\hat{f}_E&=&\frac{r^2}{l^2}-\frac{M r^2}{r^2+l_0^2}-\frac{ Q^2l_0^2}{r^2+l_0^2}- Q^2\ln(r^2+l_0^2)+{J^2 \over  4 r^2},\\ \notag
N^{\varphi}(r) &=& -{J \over 2 r^2}.
\end{eqnarray}
To obtain a rotating solution in the GB theory we closely follow
\cite{HKM2}. We have to perform the same type of boost used in the
Einstein case, but modified to account for the fact that the
higher-curvature corrections result in modifications to the AdS
length scale of the theory:
\begin{eqnarray}\label{boost} t\to \Xi_{\rm eff} t -a \varphi\,, \  \varphi \to
\frac{a t}{\ell_{\rm eff}^2} -\Xi_{\rm eff} \varphi\,,  \ \Xi_{\rm
eff}^2 = 1 + \frac{a^2}{\ell_{\rm eff}^2}\,, \end{eqnarray}
applied to our metric with \begin{eqnarray}\label{Xeff} \ell_{\rm
eff}= \sqrt{ \frac{2 \alpha}{\sqrt{X_\alpha}-1}}\,,\quad X_\alpha
= 1+\frac{4\alpha}{\ell^2}\,. \end{eqnarray} This yields
\begin{eqnarray} ds^2&=&\!-f_{GB}(\Xi_{\mbox{\tiny eff}}
dt \!-\!a d\varphi)^2\!+\!\frac{r^2}{\ell_{\mbox{\tiny eff}}^4}(adt\!-\!\Xi_{\mbox{\tiny eff}} \ell_{\mbox{\tiny eff}}^2 d\varphi)^2 \!+\!\frac{dr^2}{f_{GB}}\,,\nonumber\\
\phi& =& \ln(r/l)\,,
\label{rotbtz}
\end{eqnarray}
with $f_{GB}$ given by Eq. (16).

\section*{Acknowledgement}
KJ would like to thank Robie Hennigar for very helpful comments during the preparation if this work. The authors would like thank the anonymous reviewers for their helpful comments and constructive suggestions. 


\end{document}